# Simulations for Deep Random Secrecy Protocol

Thibault de Valroger (*)

Abstract

We present numerical simulations measuring secrecy and efficiency rate of Information Theoretically Secure protocol based on Deep Random assumption presented in former article [9]. Those simulations specifically measure the respective error rates of both legitimate partner and eavesdropper experimented during the exchange of a data flow through the protocol. The measurements of the error rates also enable us to estimate a lower bound of the Cryptologic Limit introduced in [9]. We discuss the variation of the protocol's parameters and their impact on the measured performance.

## I. Introduction and summary of former work

Modern cryptography mostly relies on mathematical problems commonly trusted as very difficult to solve, such as large integer factorization or discrete logarithm, belonging to complexity theory. No certainty exists on the actual difficulty of those problems. Some other methods, rather based on information theory, have been developed since early 90's. Those methods relies on hypothesis about the opponent (such as « memory bounded » adversary [6]) or about the communication channel (such as « independent noisy channels » [5]) ; unfortunately, if their perfect secrecy have been proven under given hypothesis, none of those hypothesis are easy to ensure in practice. At last, some other methods based on physical theories like quantum indetermination [3] or chaos generation have been described and experimented, but they are complex to implement, and, again, relies on solid but not proven and still partly understood theories.

Considering this theoretically unsatisfying situation, we have proposed in [9] to explore a new path, where proven information theoretic security can be reached, without assuming any limitation about the opponent, who is supposed to have unlimited calculation and storage power, nor about the communication channel, that is supposed to be perfectly public, accessible and equivalent for any playing party (legitimate partners and opponents). In our model of security, the legitimate partners of the protocol are using Deep Random generation to generate their secret information, and the behavior of the opponent, when inferring from public information, is governed by Deep Random assumption, that we introduce.

*(*) See contact and information about the author at last page*

## Back on the Deep random assumption

We have introduced in [9] the Deep Random assumption, based on Prior Probability theory as developed by Jaynes [7]. Deep Random assumption is an objective principle to assign probability, compatible with the symmetry principle proposed by Jaynes [7].

Before presenting the Deep Random assumption, it is needed to introduce Prior probability theory.

If we denote $\mathfrak{I}_<$ the set of all prior information available to observer regarding the probability distribution of a certain random variable $X$ ('prior' meaning before having observed any experiment of that variable), and $\mathfrak{I}_>$ any public information available regarding an experiment of $X$, it is then possible to define the set of possible distributions that are compatible with the information $\mathfrak{I} \triangleq \mathfrak{I}_< \cup \mathfrak{I}_>$ regarding an experiment of $X$; we denote this set of possible distributions as:

$$D_{\mathfrak{I}}$$

The goal of Prior probability theory is to provide tools enabling to make rigorous inference reasoning in a context of partial knowledge of probability distributions. A key idea for that purpose is to consider groups of transformation, applicable to the sample space of a random variable $X$, that do not change the global perception of the observer. In other words, for any transformation $\tau$ of such group, the observer has no information enabling him to privilege $\varphi_{\mathfrak{I}}(v) \triangleq P(X = v|\mathfrak{I})$ rather than $\varphi_{\mathfrak{I}} \circ \tau(v) = P(X = \tau(v)|\mathfrak{I})$ as the actual conditional distribution. This idea has been developed by Jaynes [7].

We will consider only finite groups of transformation, because one manipulates only discrete and bounded objects in digital communications. We define the acceptable groups $G$ as the ones fulfilling the 2 conditions below:

$(C1)$ Stability - For any distribution $\varphi_{\mathfrak{I}} \in D_{\mathfrak{I}}$, and for any transformation $\tau \in G$, then $\varphi_{\mathfrak{I}} \circ \tau \in D_{\mathfrak{I}}$

$(C2)$ Convexity - Any distribution that is invariant by action of $G$ does belong to $D_{\mathfrak{I}}$

It can be noted that the set of distributions that are invariant by action of $G$ is exactly:

$$R_{\mathfrak{I}}(G) \triangleq \left\{ \frac{1}{|G|} \sum_{\tau \in G} \varphi_{\mathfrak{I}} \circ \tau \, | \forall \varphi_{\mathfrak{I}} \in D_{\mathfrak{I}} \right\}$$

For any group $G$ of transformations applying on the sample space $F$, we denote by $\Omega_{\mathfrak{I}}(G)$ the set of all possible conditional expectations when the distribution of $X$ courses $R_{\mathfrak{I}}(G)$. In other words:

$$\Omega_{\mathfrak{I}}(G) \triangleq \{ Z(\mathfrak{I}) \triangleq E[X|\mathfrak{I}] | \forall \varphi_{\mathfrak{I}} \in R_{\mathfrak{I}}(G) \}$$

Or also:

$$\Omega_{\mathfrak{I}}(G) = \left\{ Z(\mathfrak{I}) = \int_F v \varphi_{\mathfrak{I}}(v) dv \, | \forall \varphi_{\mathfrak{I}} \in R_{\mathfrak{I}}(G) \right\}$$

The **Deep Random assumption** prescribes that, if $G \in \Gamma_{\mathfrak{I}}$, the strategy $Z_\xi$ of the opponent observer $\xi$, in order to estimate $X$ from the public information $\mathfrak{I}$, should be chosen by the opponent observer $\xi$ within the restricted set of strategies:

$$\boldsymbol{Z_\xi \in \Omega_{\mathfrak{I}}(G)} \qquad (\boldsymbol{A})$$

The Deep Random assumption can thus be seen as a way to restrict the possibilities of $\xi$ to choose his strategy in order estimate the private information $X$ from his knowledge of the public information $\mathfrak{I}$. It is a fully reasonable assumption because the assigned prior distribution should remain stable by action of a transformation that let the distribution uncertainty unchanged.

$(A)$ suggests of course that $Z_\xi$ should eventually be picked in $\bigcap_{G\in\Gamma_\mathfrak{I}}\Omega_\mathfrak{I}(G)$, but it is enough for our purpose to find at least one group of transformation with which one can apply efficiently the Deep Random assumption to the a protocol in order to measure an advantage distilled by the legitimate partners compared to the opponent.

**The security model**

We have considered in [9] secrecy protocols being specific case of the Csiszàr and Körner model, where all the information exchanged by the legitimate partners $A$ and $B$ over the main channel are fully available to the passive opponent $\xi$. $\xi$ has unlimited computing and storage power. $A$ and $B$ share initially no private information.

$\xi$ is capable to Read all published bit strings from the main channel, Store bit strings, and Make calculation on bit strings, with unlimited computing and storage power. But when $\xi$ desires to infer a private information generated by $A$'s DRG (or by $B$'s DRG) from public information, he can only do it in respect of the Deep Random assumption $(A)$ presented in previous section. This assumption creates a 'virtual' side channel for the opponent conditioning the optimal information $Z$ he can obtain to estimate the Secret Common Output information $X$.

This assumption is fully reasonable, as established in the former sections, under the condition that the DRG of $A$ and the DRG of $B$ can actually produce distributions that are truly undistinguishable and unpredictable among a set $R_\mathfrak{I}(G)$.

The protocols that we consider obey the Kerckhoffs's principle by the fact that their specifications are entirely public.

**Information Theoretically Secure Protocols (ITSP)**

The main purpose of this work is to introduce how to design an « Information Theoretically Secure Protocol » (ITSP) under Deep Random assumption, and to introduce how to generate Deep Random from classical computing resources.

We consider protocols as per the Security Model presented in previous section,

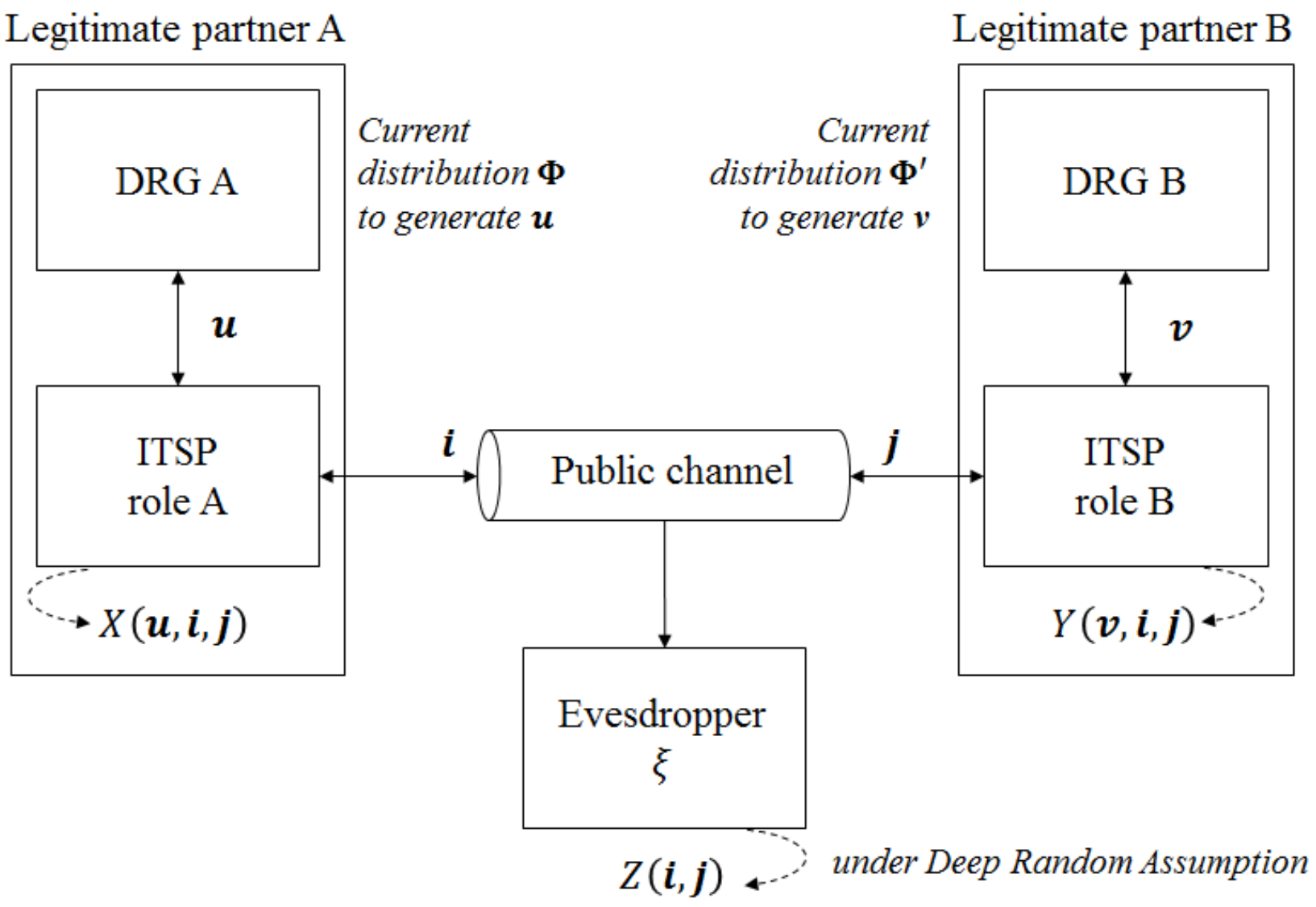

in which $A$ (resp. $B$) requests its Deep Random Generator (DRG) to obtain an experiment $u$ (resp. $v$) of a private random variable with hidden probability distribution $\Phi$ (resp. $\Phi'$); $A$ (resp. $B$) publishes the set of information $i$ (resp. $j$) on the public channel along the protocol. $A$ calculates the Secret Common Output information $X(u,i,j)$, with value in metric space $F$. $B$ calculates its estimation $Y(v,i,j)$ of the Secret Common Output information, also with value in $F$. In this model, $\mathfrak{I}_<$ is the public information available regarding the DRG of $A$ or $B$ (that are supposed to have the same design), and $\mathfrak{I}_> = \{i,j\}$ is the set of information published by the partners along the execution of the protocol.

The DRG of $A$ is run by $A$ completely privately and independently of $B$, and reversely the DRG of $B$ is run by $B$ completely privately and independently of $A$. The 2 DRG are thus not in any way secretly correlated, as one of the assumptions of the security model is that $A$ and $B$ share initially no private information.

The eavesdropping opponent $\xi$ who has a full access to the public information, calculates its own estimation $Z(\mathfrak{I})$ (that we will also shortly denote $Z(i,j)$). $Z$ is called 'strategy' of the opponent.

As introduced in the 'Deep Random assumption' section, $\mathfrak{I}_<$ designates the public information available about a DRG. We assume here that $\mathfrak{I}_<$ is the same for both DRG of $A$ and $B$, meaning that they have the same design.

From the Deep Random assumption, for any group $G$ in $\Gamma_{\mathfrak{I}}$, the set of optimal strategies for the opponent can be restricted to:

$$\Omega_{\mathfrak{I}}(G,\mathcal{P}) = \{Z(\mathfrak{I}) = E[X|\mathfrak{I}] | \forall (\varphi_{\mathfrak{I}}(u), \varphi'_{\mathfrak{I}}(v)) \in R_{\mathfrak{I}}(G) \times R_{\mathfrak{I}}(G)\}$$

Then, the protocol $\mathcal{P}$ is a ITSP if it verifies the following property $(P)$:

$$\exists \boldsymbol{G} \in \boldsymbol{\Gamma}_{\mathfrak{I}}, \boldsymbol{\alpha} > \boldsymbol{0} | \ \forall \boldsymbol{Z} \in \boldsymbol{\Omega}_{\mathfrak{I}}(\boldsymbol{G},\boldsymbol{\mathcal{P}}): \boldsymbol{H}(\boldsymbol{X}|\boldsymbol{Z}) - \boldsymbol{H}(\boldsymbol{X}|\boldsymbol{Y}) \geq \boldsymbol{\alpha} \qquad (\boldsymbol{P})$$

It has been shown ([4], [5]) that when a protocol satisfies $H(X|Z) - H(X|Y) \geq \alpha$, it can be complemented with Reconciliation and Privacy Amplification techniques to reach as close as desired from Perfect Secrecy.

We also impose a second condition for the definition of an ITSP $\mathcal{P}$. The second condition ensures that one can implement a DRG suitable for the protocol $\mathcal{P}$ thanks to a recursive and continuous generation algorithm that emulates locally the protocol. This approach is presented in section IV and introduced below. The second condition is the following:

$$\exists \boldsymbol{\alpha} > \mathbf{0} \text{ such that for any strategy } \boldsymbol{Z(\mathfrak{I})} \in \left\{E[X|\mathfrak{I}]\middle|\forall\psi(u) \in D_{\mathfrak{I}_<}\right\}\text{, there exists an actual distribution } \boldsymbol{\Psi(u)} \in D_{\mathfrak{I}_<} \text{ of the variables } X \text{ and } Y \text{ that verifies } \boldsymbol{H(X|Z) - H(X|Y) \geq \alpha} \qquad (\boldsymbol{P'})$$

The condition $(P')$ means that for any optimal strategy $Z(\mathfrak{I})$ for the distribution $\psi$, there exists a new distribution $\Psi(u) \in D_{\mathfrak{I}_<}$, such that the condition $H(X|Z) - H(X|Y) \geq \alpha$ is satisfied (we take $P_u = P_v = \Psi$ to calculate $H(X|Y)$ because symmetry in the roles $A$ and $B$ can be assumed when their DRG have same design). $\Psi$ does not depend on $\mathfrak{I}_>$ because the generation of a distribution by a DRG takes place before any instantiation of the protocol. The interest of that condition is that it enables to build a Deep Random Generator as follows:

**Deep Random Generation**

A DRG is designed in association with a given ITSP $\mathcal{P}$. The DRG executes continuously the following recursive procedure: at each step $m+1$, the generator emulates the ITSP internally and picks (through classical randomness) a new couple of identical distributions $\Phi_{m+1}(u) \in D_{\mathfrak{I}_<}, {\Phi_{m+1}}'(v) = \Phi_{m+1}(v)$ that defeats the optimal strategy (thus belonging to $\left\{E[X|\mathfrak{I}]\middle|\forall\psi(u) \in D_{\mathfrak{I}_<}\right\}$) for the past distributions for $t \leq m$. This is always possible for an ITSP as given by condition $(P')$. The source of secret entropy is the current values of the inifinite counters (several can run in parallel) of the continuous recursive process, together with the classical random that is used at each step to pick a defeating distribution.

In the present article, we will not focus on simulating a real DRG, but rather on simulating an example of ITSP, introduced in [9], working for the purpose of the simulation with a dummy emulated DRG.

**Back on the presentation of protocol $\mathcal{P}$ (introduced in [9])**

The following ITSP has been presented in [9]. In order to shortly remind the notations, the sample space of the distribution of the private information (for $A$ or $B$) is $[0,1]^n$. Considering $x = (x_1, \dots, x_n)$ and $y = (y_1, \dots, y_n)$ some parameter vectors in $[0,1]^n$ and $i = (i_1, \dots, i_n)$ and $j = (j_1, \dots, j_n)$ some Bernoulli experiment vectors in $\{0,1\}^n$, we denote :

$$x.y \text{ (resp. } i.j\text{) the scalar product of } x \text{ and } y \text{ (resp. } i \text{ and } j\text{)}$$

$$|x| \triangleq \textstyle\sum_{s=1}^{n} x_s \text{ ; } |i| \triangleq \sum_{s=1}^{n} i_s$$

$$\forall\sigma \in \mathfrak{S}_n,\ \sigma(x) \text{ represents } (x_{\sigma(1)}, \dots, x_{\sigma(n)})$$

$$\frac{x}{k} \text{ represents } \left(\frac{x_1}{k}, \dots, \frac{x_n}{k}\right) \text{ for } k \in \mathbb{R}_+^*$$

In that protocol, besides being hidden to any third party (opponent or partner), the probability distribution used by each legitimate partner also needs to have specific properties in order to prevent the opponent to efficiently evaluate $V_A$ by using internal symmetry of the distribution.

Those specific properties are :

(i) Each probability distribution $\Phi$ (for $A$ or $B$) must be « far » from its symmetric projection $\bar{\Phi}(x) \triangleq \frac{1}{n!}\sum_{\sigma\in\mathfrak{S}_n}\Phi\circ\sigma(x)$

(ii) At least one of the distribution (of $A$ or $B$) must avoid to have brutal variations (Dirac)

The technical details explaining those constraints are presented in [9]. The set of compliant distributions is denoted $\zeta(\alpha)$ where $\alpha$ is a parameter that measures the « remoteness » of a distribution from its symmetric projection.

For such a distribution $\Phi$, a tidying permutation, denoted $\sigma_\Phi$, is a specific permutation that enables to give a canonical form $\Phi\circ\sigma_\Phi$ of $\Phi$, such form being useful to « synchronize » two distributions by transitivity. Again, technical details are given in [9]. One can just say here that it is linked to the quadratic matrix whose coefficient is $M_\Phi(u,v) \triangleq \int_{[0,1]^n} x_u x_v \Phi(x)dx$, by minimizing

$$\sum_{u,v\in I_0\times\bar{I}_0} M_{\Phi\circ\sigma_\Phi}(u,v) = \min_{\sigma\in\mathfrak{S}_n}\left(\sum_{u,v\in I_0\times\bar{I}_0} M_{\Phi\circ\sigma}(u,v)\right)$$

where $I_0 \triangleq \{1,\dots,{}^{n}/_{2}\}$.

Here are the steps of the proposed protocol:

$A$ and $B$ are two AE, called the legitimate partners. The steps of the protocol $\mathcal{P}(\alpha,n,k,L)$ are the followings:

*Step 1 – Deep Random Generation: A and B pick independently the respective probability distributions $\Phi$ and $\Phi' \in \zeta(\alpha)$, so that $\Phi$ (resp. $\Phi'$) is secret (under Deep Random assumption) for any observer other than A (resp. B) beholding all the published information. A draws the parameter vector $x \in [0,1]^n$ from $\Phi$. B draws the parameter vector $y \in [0,1]^n$ from $\Phi'$.*

*Step 2 – Degradation: A generates a Bernoulli experiment vectors $i \in \{0,1\}^n$ from the parameter vector $\frac{x}{k}$. A publishes $i$. B generates a Bernoulli experiment vectors $j \in \{0,1\}^n$ from the parameter vector $\frac{y}{k}$. B publishes $j$.*

*Step 3 – Dispersion: A and B also pick respectively a second probability distribution $\Psi$ and $\Psi' \in \zeta(\alpha)$ such that it is also secret (under Deep Random assumption) for any observer other than A (resp. B). $\Psi$ is selected also such that $\int_{|x|\in[k|i|-\sqrt{n},k|i|+\sqrt{n}]}\Psi(x)dx \geq \frac{1}{2\sqrt{n}}$ in order to ensure that $|i|$ is not an unlikely value for $\approx \left|\frac{x}{k}\right|$ (same for $\Psi'$ by replacing $x$ by $y$ and $i$ by $j$). $\Psi$ (resp. $\Psi'$) is used to scramble the publication of the tidying permutation of A (resp. B). A (resp. B) calculates a permutation $\sigma_d[i]$ (resp. $\sigma'_d[j]$) representing the reverse of the most likely tidying permutation on $\Psi$ (resp. $\Psi'$) to produce i (resp. j). In other words, with $i$, $\sigma_d[i]$ realizes :*

$$\max_{\sigma\in\mathfrak{S}_n}\int_x P(i|x)\Psi\circ\sigma_\Psi\circ\sigma^{-1}(x)dx$$

*Then $A$ (resp. B) draws a boolean $b \in \{0,1\}$ (resp. $b'$) and publishes in a random order $(\mu_1, \mu_2) = t^b(\sigma_d[i], \sigma_\Phi)$, (resp. $(\mu'_1, \mu'_2) = t^{b'}(\sigma'_d[j], \sigma_{\Phi'})$) where $t$ represents the transposition of elements in a couple.*

*Step 4 – Synchronization: $A$ (resp. B) chooses randomly $\sigma_A$ (resp. $\sigma_B$) among $(\mu'_1, \mu'_2)$ (resp. $(\mu_1, \mu_2)$).*

*Step 5 – Decorrelation: $A$ computes $V_A = \frac{\sigma_\Phi{}^{-1}(x).\sigma_A{}^{-1}(j)}{n}$, $B$ computes $V_B = \frac{\sigma_B{}^{-1}(i).\sigma_{\Phi'}{}^{-1}(y)}{n}$. $V_A$ and $V_B$ are then transformed respectively by $A$ and $B$ in binary output thanks to a sampling method described hereafter. At this stage the protocol can then be seen as a broadcast model with 2 Binary Symmetric Channels (BSC), one between $A$ and $B$ and one between $A$ and $\xi$ who computes a certain $V_\xi$, called $\xi$'s strategy, that is to be transformed in binary output by the same sampling method than for $A$ and $B$. It is shown in Theorem 1 of [9] that those 2 BSC are partially independent, which enable to create Advantage Distillation as shown in [5].*

*Step 5' – Advantage Distillation: by applying error correcting techniques with code words of length L between $A$ and $B$, as introduced in [5], we show in Theorem 1 of [9] that we can then create advantage for $B$ compared to $\xi$ in the error rates of the binary flows resulting from the error correcting code.*

*Step 6: classical Information Reconciliation and Privacy Amplification (IRPA) techniques then lead to get accuracy as close as desired from perfection between estimations of legitimate partners, and knowledge as close as desired from zero by any unlimitedly powered opponent, as shown in [4].*

The choices of the parameters $(\alpha, n, k, L)$ are theoretically discussed in proof of main Theorem in [9]. They are discussed based on practicle simulations in the following section. They are set to make steps 5, 5' and 6 possible.

The Degradation transformations $x \longmapsto \frac{x}{k}$ and $y \longmapsto \frac{y}{k}$ with $k > 1$ at step 2 are the ones that prevent the use of direct inference by the opponent, and of course, the Deep Random Generation at step 1 prevents the use of Bayesian inference based on the knowledge of the probability distribution. The synchronization step 4 is designed to overcome the independence between the choices of the distributions of $A$ and $B$, and needs that the distributions to have special properties ($\in \zeta(\alpha)$) in order to efficiently play their role. It is efficient in $1/4$ of cases (when $B$ picks $\sigma_B = \sigma_\Phi$ and $A$ picks $\sigma_A = \sigma_{\Phi'}$, which we will call favorable cases). And to prevent $\xi$ from gaining knowledge of $\sigma_\Phi$, Dispersion step 3 mixes $\sigma_\Phi$ within $(\mu_1, \mu_2)$ with another permutation $\sigma_d[i]$ (and $\sigma_{\Phi'}$ within $(\mu'_1, \mu'_2)$ with another permutation $\sigma'_d[j]$) that (1) is undistinguishable from $\sigma_\Phi$ knowing $i$, and (2) manages to make the estimation of $\xi$ unefficient as shown in [9]. We denote the following set of strategies (invariant by transposition of $(\mu_1, \mu_2)$ or $(\mu'_1, \mu'_2)$):

$$\Omega'_\# \triangleq \left\{\omega(i, j, (\mu_1, \mu_2), (\mu'_1, \mu'_2)) \,\middle|\, \forall b, b' \in \{0,1\}: \omega\left(i, j, \tau^b(\mu_1, \mu_2), \tau^{b'}(\mu'_1, \mu'_2)\right) = \omega(i, j, (\mu_1, \mu_2), (\mu'_1, \mu'_2))\right\}$$

Because of Deep Random assumption $(A)$ over the group $\{\mathrm{Id}, t\}$ applied to the distribution of $(\mu_1, \mu_2)$ and $(\mu'_1, \mu'_2)$, the strategy of the opponent can thus be restricted to $V_\xi \in \Omega'_\#$.

$i$ is entirely determined by $|i|$ and a permutation, which explains the constraint and transformation applied on $\Psi$ in step 3 to make $\sigma_\Phi$ and $\sigma_d[i]$ indisguishable knowing $i$ (same with ${\sigma'}_d[j]$, $\sigma_{\Phi'}$, and $j$).

The synchronization step has a cost when considering the favorable cases: $\xi$ knows that $\Phi$ and $\Phi'$ are synchronized in favorable cases, which means in other words that $\xi$ knows that an optimal (or quasi optimal) permutation is applied to $\Phi'$. This also means that in favorable cases, all happen like if when A picks $\Phi \circ \sigma$ instead of $\Phi$, the result of the synchronization is that $B$ uses $\Phi' \circ \sigma$ instead of $\Phi'$. Starting from the most general strategy $\omega \in \Omega'_\#$ for $\xi$, we also consider the following additional restrictions applicable to the favorable cases:

- Restriction to the strategies of the form $\omega(i,j)$, because $(\sigma_d[i], {\sigma'}_d[j])$ depends only on $(i,j)$ and not on $\Phi$ neither $\Phi'$,
- And then restriction to the set of strategies such that $\omega_{i,j} = \omega_{\sigma(i),\sigma(j)}, \forall \sigma \in \mathfrak{S}_n$, in other words strategies invariant by common permutation on $i, j$.

which leads to define the more restricted set of strategies:

$$\Omega_\#(G,\mathcal{P}) = \left\{\omega \in [0,1]^{2^{2n}} | \omega(\sigma(i),\sigma(j),\mu_1,\mu_2,{\mu'}_1,{\mu'}_2) = \omega(\sigma(i),\sigma(j)) = \omega(i,j)\ , \forall \sigma \in \mathfrak{S}_n\right\}$$

The step 5 is called Decorrelation because at this step, thanks to the Deep Random Assumption, we have managed to create a protocol that can be equivalently modelized by a broadcast communication over 2 partially independent (not fully correlated) BSC, as shown in the main Theorem in [9], and also that consequently, it is possible to apply error correcting techniques to create Advantage Distillation as established in [5].

We can use the following very basic technique to transform $V_A$ and $V_B$ in an intermediate binary flow as introduced in step 5: we can typically sample a value in $[0,1]$ like $V_A$ or $V_B$ with a gauge being a multiple of the variance $E\left[\left(V_{A|\sigma_A=\sigma_{\Phi'}} - V_{B|\sigma_B=\sigma_\Phi}\right)^2\right]^{1/2} = O\left(\frac{1}{\sqrt{nk}}\right)$. The multiplicative factor $K$ is chosen such that :

$$\frac{1}{\sqrt{nk}} \ll \frac{K}{\sqrt{nk}} \ll \frac{1}{\sqrt{n}}$$

and therefore, each experiment of the protocol can lead respectively $A$ and $B$ to distill an intermediate digit defined by :

$$\widetilde{e_A} = \left\lceil \frac{V_A\sqrt{nk}}{K} \right\rceil \bmod 2,\ \widetilde{e_B} = \left\lceil \frac{V_B\sqrt{nk}}{K} \right\rceil \bmod 2$$

Regarding the legitimate partners, when $B$ picks $\sigma_B = \sigma_\Phi$ and $A$ picks $\sigma_A = \sigma_{\Phi'}$, the choice of $\sigma_A$ and $\sigma_B$ remain independent from $i, j$, so that $i$ and $j$ remain draws of independent Bernoulli random variables, then allowing to apply Chernoff-style bounds for the legitimate partners. When $B$ picks $\sigma_B = \sigma_d[i]$ or $A$ picks $\sigma_A = {\sigma'}_d[j]$, this is no longer true and $V_B$ or $V_A$ become erratic, which will lead to error detection by error correcting code at step 5'.

The heuristic table analysis of the protocol is then the following:

| | $B$ picks $\sigma_B = \sigma_\Phi$ among $(\mu_1, \mu_2)$ | $B$ picks $\sigma_B = \sigma_d[i]$ among $(\mu_1, \mu_2)$ |
|---|---|---|
| $A$ picks $\sigma_A = \sigma_{\Phi'}$ among $({\mu'}_1, {\mu'}_2)$ | $A$ and $B$ respective estimations are close in ~100% of cases, and thus | $A$ and $B$ respective estimations are not close which leads to error |
| | both obtain accurate estimation of the combined shared secret in ~100% of cases.<br><br>$\xi$ cannot make accurate estimation of the combined shared secret in at least ~25% of cases (if $\xi$ tries to have a strategy depending on $(\mu_1, \mu_2, \mu'_1, \mu'_2)$, then $(\sigma_d[i], \sigma'_d[j])$ is indistinguishable from $(\sigma_\Phi, \sigma_{\Phi'})$ and is thus picked by $\xi$ in 25% of cases. | detection and finally discarding. |
| $A$ picks $\sigma_A = \sigma'_d[j]$ among $(\mu'_1, \mu'_2)$ | $A$ and $B$ respective estimations are not close which leads to error detection and finally discarding. | $A$ and $B$ respective estimations are not close which leads to error detection and finally discarding. |

This is a heuristic reasoning, and we must rather consider most general strategies $\omega(i, j, \mu_1, \mu_2, \mu'_1, \mu'_2)$ and write the probability equations with the appropriate group transform, under Deep Random assumption, which is done in [9]. But this little array explains why we create partial independence between the BSC and consequently then an advantage for the legitimate partners compared to the opponent, bearing in mind that $(\mu_1, \mu_2)$ (resp. $(\mu'_1, \mu'_2)$) are absolutely undistinguishable knowing $i$ (resp. $j$), due to the fact that the distributions $\Phi$ and $\Psi$ (resp. $\Phi'$ and $\Psi'$) are unknown and thus also absolutely undistinguishable by $\xi$.

The sampling method presented above has the drawback of the border effect. If the reference value $V_B$ is too close from one of the sampling frontier $\left\{\frac{tK}{\sqrt{nk}}\right\}_{t\in\mathbb{N}}$, then the sampling process becomes unefficient. In order to avoid the border effect, one can bring a little improvement to the protocol by allowing $B$ to publish :

$$\rho_B = V_B - \frac{K}{\sqrt{nk}}\left\lfloor V_B \frac{\sqrt{nk}}{K} \right\rfloor$$

and then to replace $V_B$ by $V_B \to V_B + \frac{K}{2\sqrt{nk}} - \rho_B$ in order to center $V_B$ within the sampling comb. This of course results in applying the same transform on $V_A$ and $\omega$ :

$$V_A = \frac{K}{\sqrt{nk}}\left\lfloor \left(V_A + \frac{K}{2\sqrt{nk}} - \rho_B\right)\frac{\sqrt{nk}}{K} \right\rfloor + \frac{K}{2\sqrt{nk}}$$

$$\omega = \frac{K}{\sqrt{nk}}\left\lfloor \left(\omega + \frac{K}{2\sqrt{nk}} - \rho_B\right)\frac{\sqrt{nk}}{K} \right\rfloor + \frac{K}{2\sqrt{nk}}$$

The publication of $\rho_B$ does not bring any additional information to the opponent regarding the valuable secret information being the parity of $\left\lfloor V_B \frac{\sqrt{nk}}{K} \right\rfloor$.

The following section presents simulation and results confirming numerically that one can obtain strictly positive values of the Cryptologic Limit introduced in [9] with appropriate values of $(n, k, K, L)$. The induced lower bound of the Cryptologic Limit is also measured.

## II. Presentation of simulation and results

### Handling dummy DRGs

In our simulation, we don't focus on DRG but specifically on the ITSP presented in the previous section. The DRG of each partner is emulated from a fixed distribution $\Phi_0$ in $\zeta(\alpha)$ with $\sigma_{\Phi_0} = \mathrm{Id}_{\mathfrak{S}_n}$.

$\Phi_0$ is defined by :

$$P\left(\sum_{s=1}^{n/2} x_s = t\right) = P\left(\sum_{s=n/2+1}^{n} x_s = t'\right) = \frac{2}{n}, \forall t, t' \in \{0, \dots, n/2\}$$

Its quadratic matrix is:

$$\left(\begin{array}{ccc|ccc} 1/2 & \cdots & 1/3 & 1/4 & \cdots & 1/4 \\ \vdots & \ddots & \vdots & \vdots & & \vdots \\ 1/3 & \cdots & 1/2 & 1/4 & \cdots & 1/4 \\ \hline 1/4 & \cdots & 1/4 & 1/2 & \cdots & 1/3 \\ \vdots & & \vdots & \vdots & \ddots & \vdots \\ 1/4 & \cdots & 1/4 & 1/3 & \cdots & 1/2 \end{array}\right)$$

And it verifies, still with notations of [9] :

$$\left\| M_{\Phi_0} - \overline{M_{\Phi_0}} \right\|_c = \sqrt{\alpha} = \frac{1}{12} + O\left(\frac{1}{n}\right)$$

### Input parameters, dispersion and error correcting

The simulation is executed for a given set of parameters $(n, k, K, L)$ and, at each execution, it repeats $R$ times the protocol. The dispersion is simulated by the generation of $\sigma_d[i]$ (resp. $\sigma'_d[j]$) in such a way that $i \subset \sigma_d[i](I_0)$ (resp. $j \subset \sigma'_d[j](I_0)$) and randomly, one time over two in average, $A$ (resp. also and independently for $B$) chooses $\sigma_A = \sigma'_d[j]$ rather than $\sigma_A = \mathrm{Id}$ (resp. $\sigma_B = \sigma_d[i]$ rather than $\sigma_B = \mathrm{Id}$), which simulates the random choice of $A$ (resp. also and independently for $B$) among $(\mu'_1, \mu'_2)$ (resp. $(\mu_1, \mu_2)$). One can notice that it is useless to introduce random permutation in the draw of $\Phi_0$ because it is easy to verify that $\sigma_d[i]_{[\Phi_0 \circ \mu]} = \mu^{-1} \circ \sigma_d[i]_{[\Phi_0]}$ and thus

$$\Phi_0 \circ \mu \circ \sigma_d[i]_{[\Phi_0 \circ \mu]} = \Phi_0 \circ \sigma_d[i]_{[\Phi_0]}$$

In those simulations, we mainly focus on Advantage Distillation. The effect of Privacy Amplification will be discussed for one of the simulations. Information Reconciliation is not fully considered, but error correcting methods need to be handled at step 5' in order to achieve Advantage Distillation. We will use the (non-optimal) method proposed by Maurer in [5]; the codewords $v_A$ chosen by $A$ can only be $(0,0,\dots,0)_L$ or $(1,1,\dots,1)_L$ depending on $e_A = 0$ or $e_A = 1$. $B$ publicly discards all decoded sequence $v_B$ that is not $(0,0,\dots,0)_L$ or $(1,1,\dots,1)_L$ and obviously decodes accordingly $e_B = 0$ if $|v_B| = 0$, and $e_B = 1$ if $|v_B| = L$.

The discarding rate is denoted $D$, corresponding to the average ratio of discarded sequences.

## Behavior of the opponent

The behavior of the opponent is tested with 3 canonical strategies among $\Omega'_{\#}(G,\mathcal{P})$ :

$$\omega_0(i,j) = \frac{ki\cdot j}{n}$$

$$\omega_1(i,j) = \frac{k|i||j|}{n^2}$$

$$\omega_2(i,j,(\mu_1,\mu_2),(\mu'_1,\mu'_2)) = \frac{2k\left(\left(\sum_{r\in\sigma_{\xi,B}(I_0)} i_r\right)\left(\sum_{r\in\sigma_{\xi,A}(I_0)} j_r\right) + \left(\sum_{r\in\overline{\sigma_{\xi,B}(I_0)}} i_r\right)\left(\sum_{r\in\overline{\sigma_{\xi,A}(I_0)}} j_r\right)\right)}{n^2}$$

where $\sigma_{\xi,B}$ is a permutation randomly chosen by $\xi$ among $(\mu_1,\mu_2)$ and $\sigma_{\xi,A}$ is a permutation randomly chosen by $\xi$ among $(\mu'_1,\mu'_2)$.

A pessimistic approach is adopted in which, the best of the three is selected for the opponent at the end of the execution of a simulation, separately for each measured criteria, when delivering the result (« best » meaning the one that gives the most favorable criteria for the opponent, typically the higher opponent's knowledge rate over the private information, or the lesser Cryptologic Limit).

Those three strategies are chosen on purpose:

- $\omega_0$ represents the only strategy that equalizes $E[\omega_0|(x,y)\&(\sigma_B=\sigma_{\Phi\circ\sigma})] = E[V_B|(x,y)\&(\sigma_B=\sigma_{\Phi\circ\sigma})]$ (See [9] section II for the details) ;
- $\omega_1$ represents the canonical strategy beating the partners in case their distributions are not synchronized (e.g. which needs to avoid $\sigma_B=\sigma_\Phi$ and $\sigma_A=\sigma_{\Phi'}$).
- $\omega_2$ represents the canonical strategy when $\xi$ tries to « guess » and use the permutations $\sigma_\Phi$ and $\sigma_{\Phi'}$ ; $\omega_2$ is efficient when $\xi$ guess correctly.

## Output of simulation

Each execution of the simulation outputs 3 information:

(i) the error rate for the legitimate partners – denoted $\varepsilon$,
(ii) the knowledge rate for the opponent – denoted $\varepsilon'$, and
(iii) a lower bound of Cryptologic Limit – denoted $CL$.

$P(e_A\neq e_B)$ and $\left|P(e_E=e_B)-\frac{1}{2}\right|$ can be seen respectively as the error rate and 'eavesdropping' rate, so both with values in [0,1]. We decided in [9], by convention that fits the common sense, that $P(e_A\neq e_B)$ cannot be greater than $1/2$, and $P(e_E=e_B)$ cannot be lower than $1/2$ ; indeed, if $P(e_A\neq e_B)>1/2$ one can replace $e_B$ by $\overline{e_B}$ (we then denote $e_B{}^*$ the final choice for $e_B$), and if $P(e_E=e_B)<1/2$ one can replace $e_E$ by $\overline{e_E}$ (we then denote $e_E{}^*$ the final choice for $e_E$). And therefore, we can define $\varepsilon$ as the net error rate, and $\varepsilon'$ as the net knowledge rate (by the opponent), with the following formulas:

(i) $\varepsilon = 2\left(\min\left(P(e_A\neq e_B),1-P(e_A\neq e_B)\right)\right) = 2P(e_A\neq e_B{}^*)$

(ii) $\varepsilon' = 2\left(\max\left(P(e_E=e_B),1-P(e_E=e_B)\right)-\frac{1}{2}\right) = 2\left(P(e_A=e_E{}^*)-\frac{1}{2}\right) > 0$

$CL$ as introduced in [9], represents the rate of perfectly reliable bit transmitted over the total amount of bit transmitted, which in the case of the protocol is :

(iii) $CL = (1-D)\frac{1-\varepsilon-\varepsilon\prime}{6nL}$ (one can remark that it is possible to only transmit $\mu_x(I_0)$ instead of $\mu_x$, which can be coded with $n$ bits instead of $n \log_2 n$).

The simulation also uses the border effect improvement presented in section I, unless the opposite choice is explicitly mentioned in the condition of the simulation.

The 3 output information are measured as averaged over all the $R$ repetitions of the protocol within an execution of the simulation.

## Presentation of the results

We have executed series of simulations with 3 dimensional coursing of parameters $(n, k, K)$, and with $R$ being equal to at least 5,000.

The simulation program is written in C language, generated with Visual Studio 2015 on Windows 7 - 64 bits. It uses the random generator RAND( ) of the stdlib.h standard library.

In the following results, the sampling parameter $K$ is expressed as a multiple of $\frac{1}{2\sqrt{nk}}$.

In all the graphics below, the abscissa represents the coursing parameter values, the left ordinate represents the values of error rate $\varepsilon$ and knowledge rate $\varepsilon'$, the right ordinate represents the values of the Cryptologic Limit's lower bound $CL$ with value being multiple of $10^{-6}$, and the title of the graphic gives the values of the fixed parameters, and the range of the coursing parameter.

In the first graphic presented below, the values of the fixed parameters are $(n = 10000,\ k = 16,\ L = 4)$, $K$ is coursing from 1 to 10.

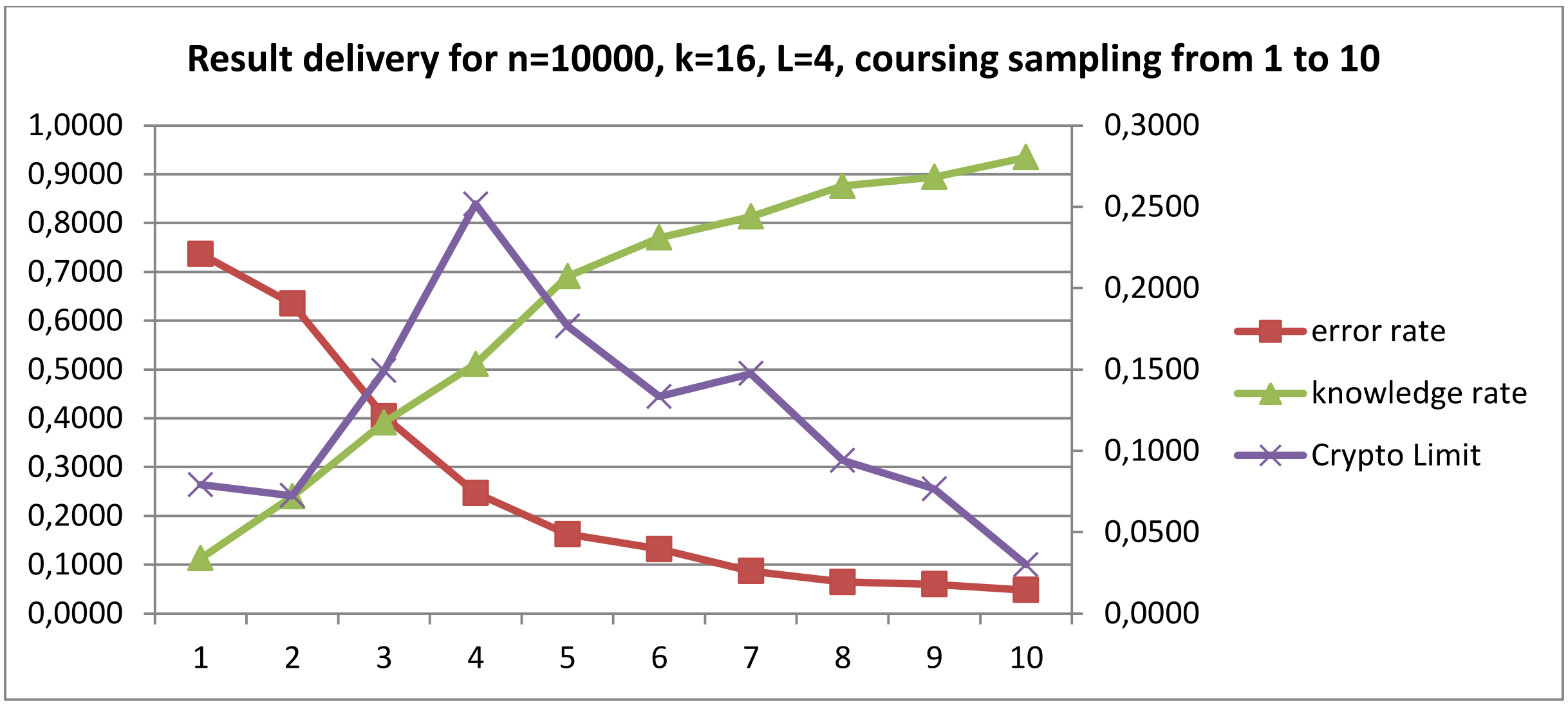

The maximum of $CL$ is obtained for $K = 4$. As $K$ is increasing, the error rate is decreasing, and the knowledge rate is increasing, as expected.

The graphic below performs on the same parameters' values but without the border effect avoidance improvement. One can see that $CL$ reaches a lesser maximum.

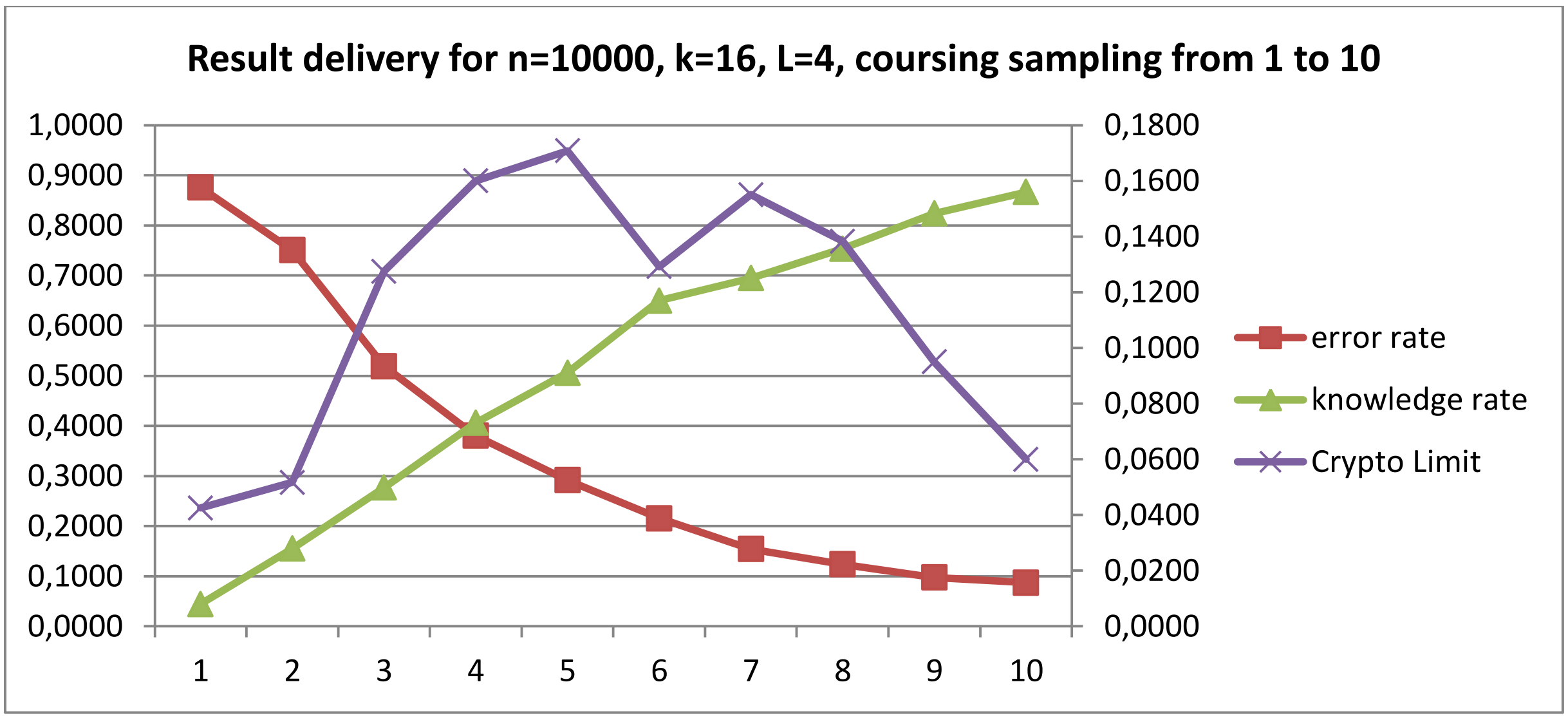

The second graphic below illustrates the use of a larger value for $k = 40$. One can see that it enables to reach a lower value of the error rate but the price to pay is a quicker increase of the knowledge rate, eventually resulting in a lower maximum of $CL$. Indeed, increasing $k$ too much results in lowering the sharpness of the synchronization process, and then favoring the efficiency of opponent's strategy $\omega_1$.

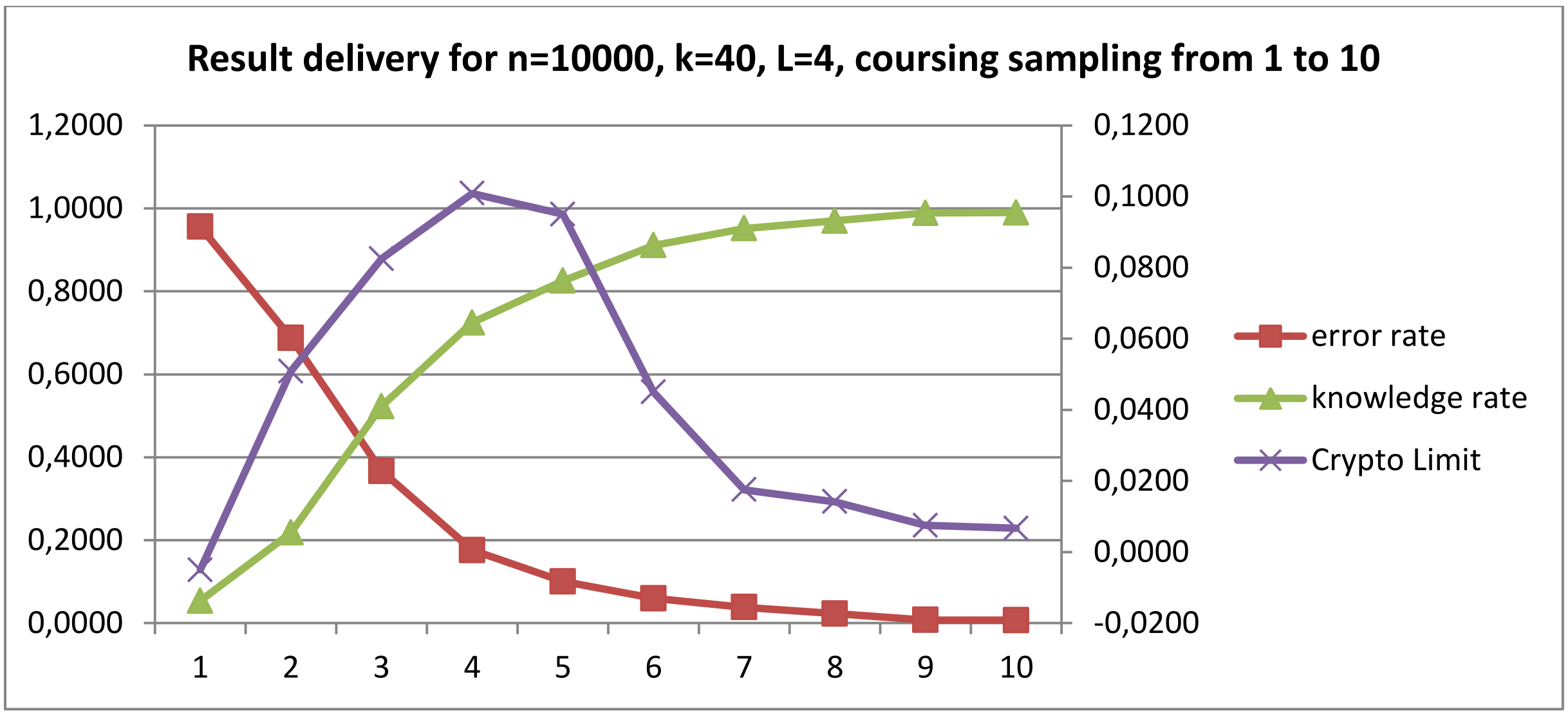

The third graphic below represents the coursing of parameter $k$ when $(n = 10000, \; K = 4, \; L = 4)$ are fixed. It illustrates that, from a certain point, increasing $k$ has no positive effect on the outputs. The maximum of $CL$ coincides with the optimal lowest value of the knowledge rate. The increase of $k$ has a strong effect on lowering the knowledge rate at the begining, but then, as already discussed, creates a back effect that lowers the efficiency of synchronization.

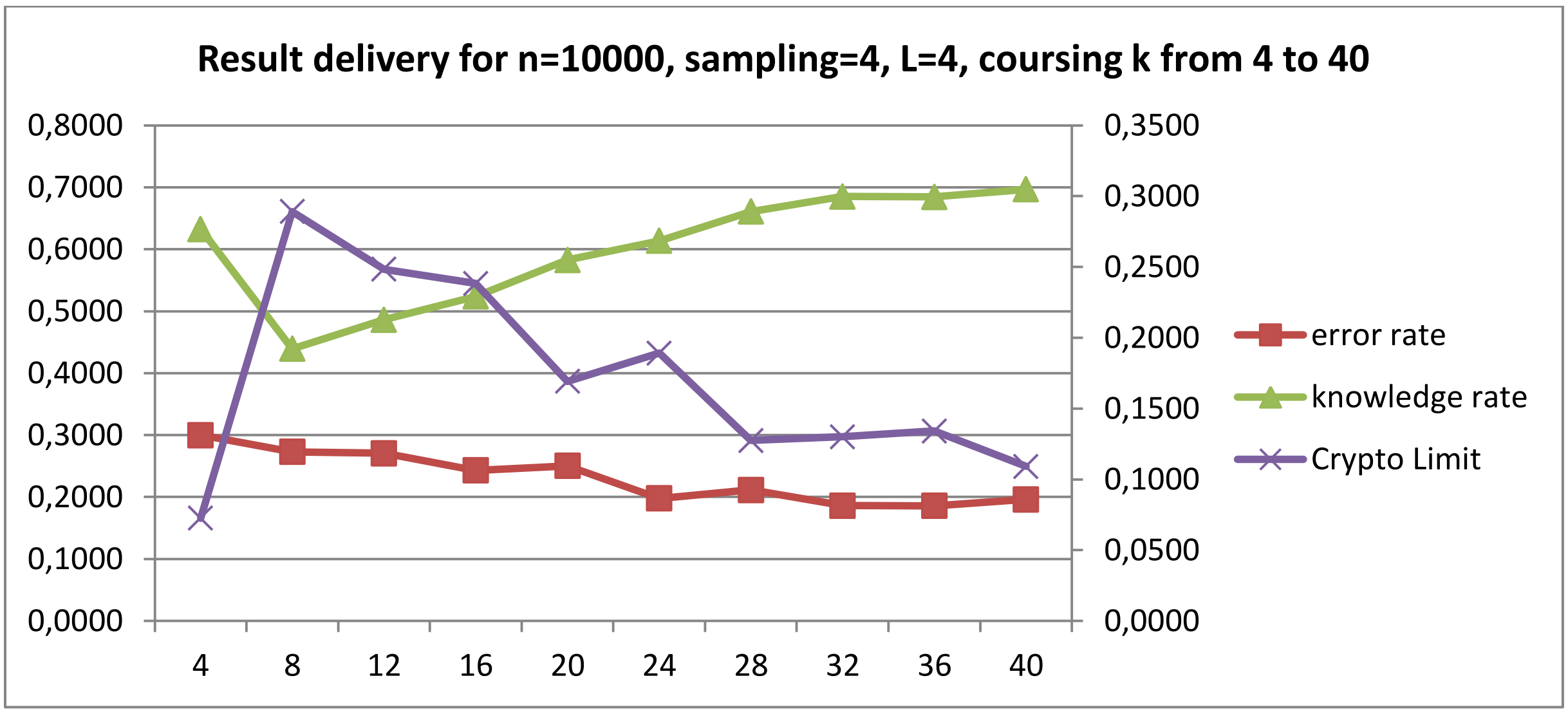

The forth graphic below presents the effect of increasing the codelength $L$. The fixed parameters are $(n = 10000,\ k = 16,\ K = 4)$ ; the codelength is coursing from 1 to 10.

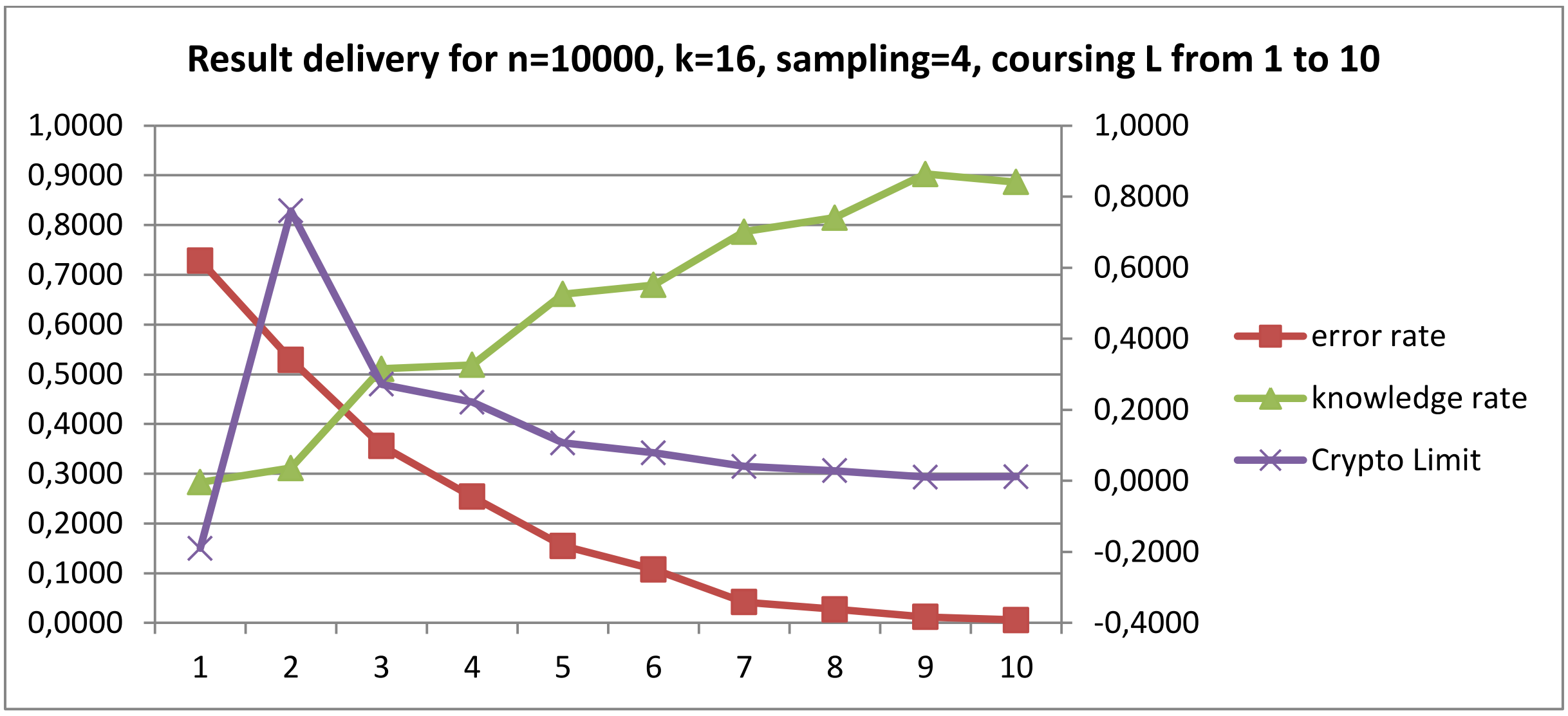

The maximum value for $CL$ is rapidly obtained for $L = 2$. As expected when $L$ increases, the error rate decreases and knowledge increases.

The fifth graphic below presents a larger value of parameter $n = 90000$. The other fixed parameters are $(K = 4, L = 4)$. The learning from this graphic is that increasing the value of $n$ does not really help, even for the values of the error and knowledge rates, compared to the best results obtained with typically $n = 10000$. It means that the internal parameters of the protocol are sufficient to create Advantage Distillation, but are not efficient enough to create strong Information Reconciliation and Privacy Amplification (IRPA). This suggests to add external sophisticated method for IRPA, as proposed in [4] and [13] to reach very low values of $\varepsilon$ and $\varepsilon'$ acceptable for real implementation. We will discuss this in further work.

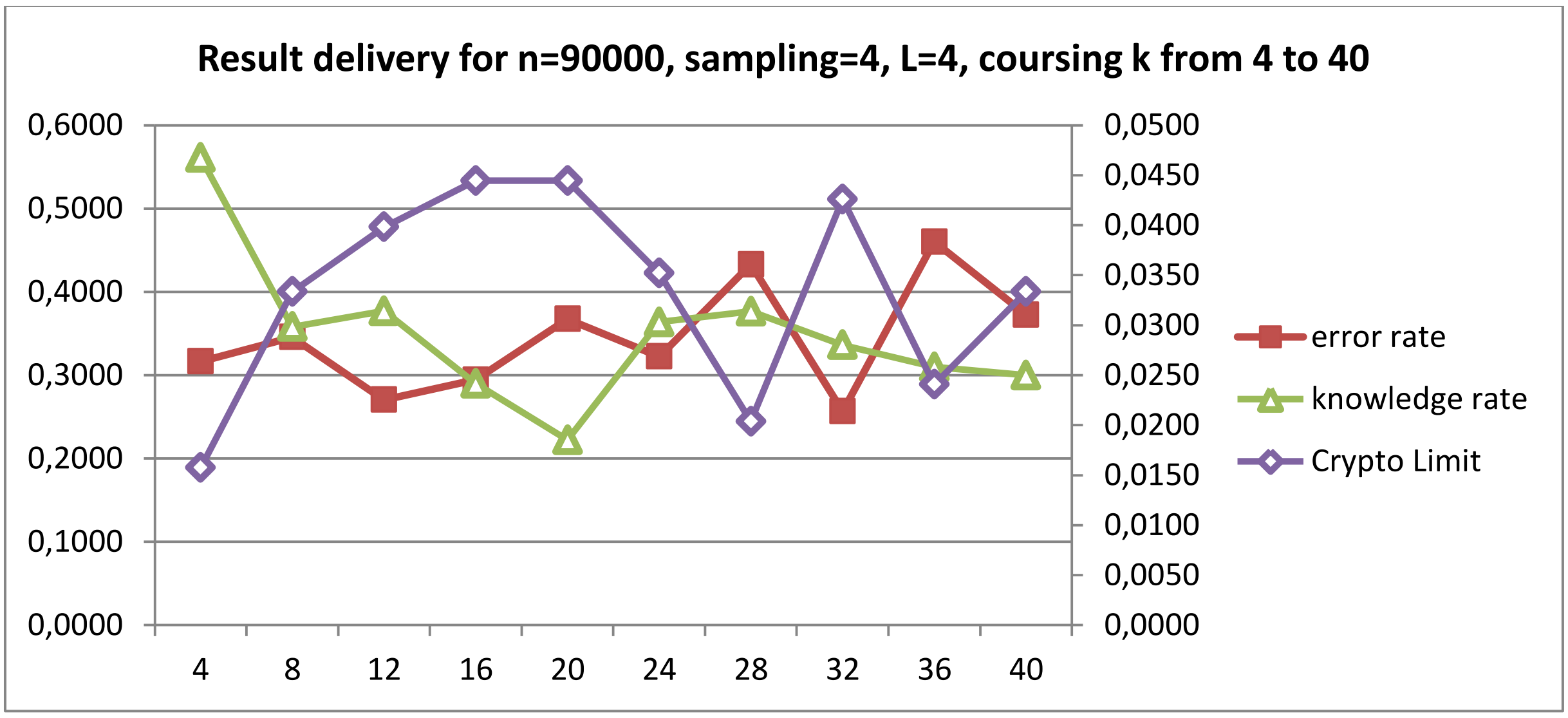

Finally, in the sixth graphic, we present the effect of Privacy Amplification. We start from the result obtained in the fourth graphic for the highest value of $L$, which gives the fixed parameters ($n = 10000,\ k = 16,\ K = 4,\ L = 10$). We apply a simple Privacy Amplification method corresponding to the universal hashing function « multiply-add-shift » $h_{a,b}(x) = \left((ax + b) \bmod 2^{PA}\right) \operatorname{div} 2^{PA-1}$, for which parameters $a, b$ are picked for each new output digit (distilled from sequences of $PA$ input digits). The codewords size $PA$ varies from 1 to 30 :

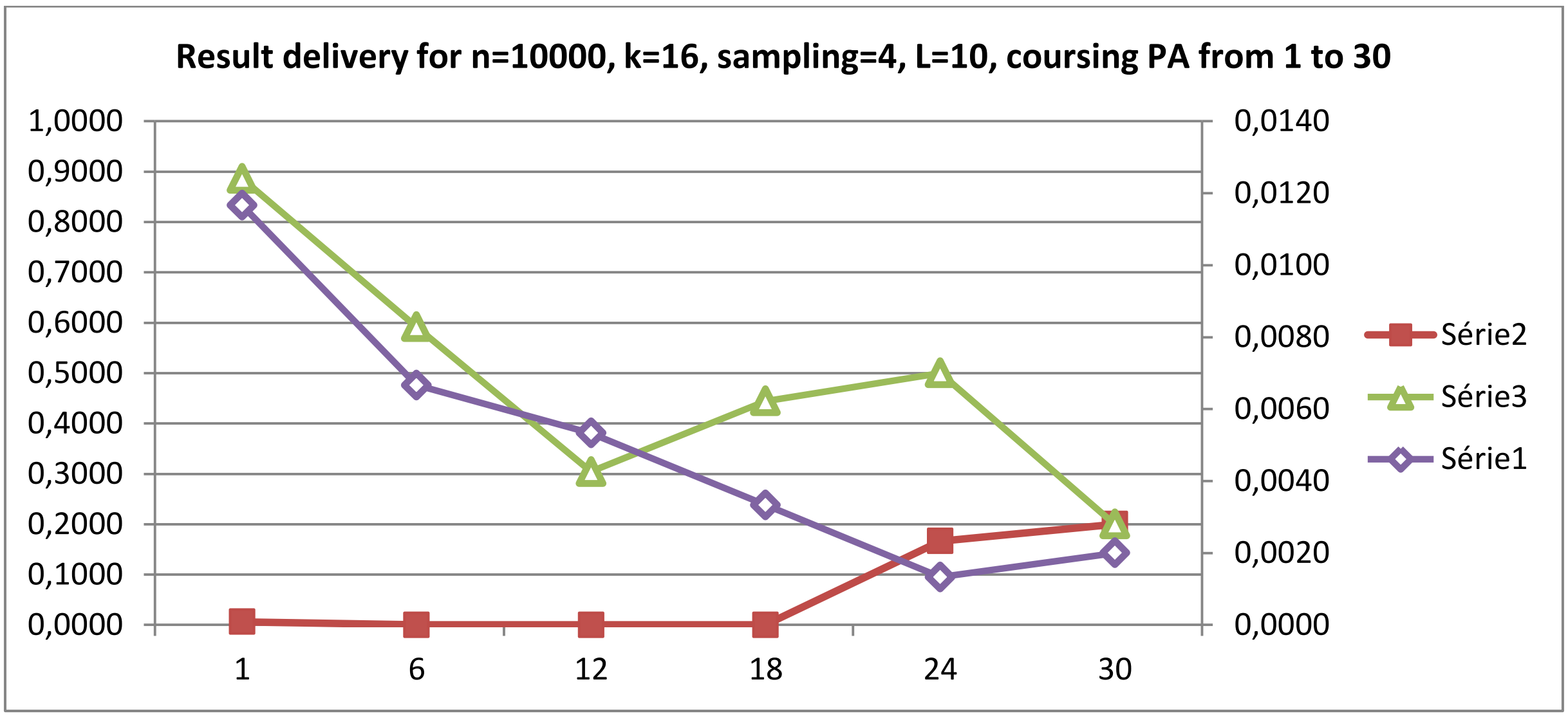

The reduction of the knowledge rate is of course at cost of the increase of the error rate, and at the overall decrease of $CL$. It is stated however without proof that IRPA techniques are more efficient than increasing the value of $n$.

As a main conclusion of this work, it is obtained that the Cryptologic Limit $C$ introduced in [9] can be approximately lower bounded by:

$$C \gtrsim 0{,}76 \cdot 10^{-6}$$

as per result of the fourth simulation (fourth graphic) presented above. That lower bound is absolutely not considered as optimal, and should be considered with the remarks that (i) we did not use real DRG but just emulations, (ii) we only considered the 3 strategies $(\omega_0, \omega_1, \omega_2)$ for the opponent. However, (i) the choice of $(\omega_0, \omega_1, \omega_2)$ appear as the most natural, and (ii) we gave an additional artificial advantage for the opponent by systematically choosing the best strategy among $(\omega_0, \omega_1, \omega_2)$ from the measures, which an actual opponent would not be able to do.

**Who is the author ?**

I have been an engineer in computer science for 20 years. My professional activities in private sector are related to IT Security and digital trust, but have no relation with my personal research activity in the domain of cryptology. If you are interested in the topics introduced in this article, please feel free to establish first contact at tdevalroger@gmail.com